\documentclass[aps,prd,twocolumn,showpacs,superscriptaddress,groupedaddress,nofootinbib]{revtex4}  
\usepackage{graphicx}  
\usepackage{dcolumn}   
\usepackage{bm}        
\usepackage{amsmath,amssymb,amsthm,amstext,amsfonts}   

\usepackage[english]{babel}
\usepackage[utf8]{inputenc}  
\usepackage{hyperref} 
\usepackage{natbib}

\newcommand{\beq}{\begin{equation}}
\newcommand{\eeq}{\end{equation}}
\newcommand{\bey}{\begin{eqnarray}}
\newcommand{\eey}{\end{eqnarray}}
\newcommand{\bal}{\begin{align}}
\newcommand{\eal}{\end{align}}

\newtheorem*{assumption1}{Law 1}
\newtheorem*{assumption2}{Law 2}
\newtheorem*{assumption2p}{Law 2}
\newtheorem*{assumption3}{Law 3}

\begin{document}



\title{Notes on several phenomenological laws of quantum gravity}
\author{Jean-Philippe Bruneton}
\affiliation{Namur Center for Complex systems (naXys),\\ University of Namur, Belgium}
\email{jpbruneton@gmail.com}       
\date{\today}

\begin{abstract}
\noindent Phenomenological approaches to quantum gravity try to infer model-independent laws by analyzing thought experiments and combining both quantum, relativistic, and gravitational ingredients. We first review these ingredients -three basic inequalities- and discuss their relationships with the nature of fundamental constants. In particular, we argue for a covariant mass bound conjecture: in a spacetime free of horizon, the mass inside a surface $A$ cannot exceed $16 \pi G^2 m^2< A $, while the reverse holds in a spacetime with horizons. This is given a precise definition using the formalism of light-sheets. We show that $\hbar/c$ may be also given a geometrical interpretation, namely $4 \pi \hbar^2/m^2< A$.\\
We then combine these inequalities and find/review the following: (1) Any system must have a size greater than the Planck length, in the sense that there exists a minimal area (2) We comment on the Minimal Length Scenarios and the fate of Lorentz symmetry near the Planck scale (3) Quanta with transplanckian frequencies are allowed in a large enough boxes (4) There exists a mass-dependent maximal acceleration given by $m c^3/\hbar$ if $m<m_p$ and by $c^4/G m$ if $m>m_p$ (5) There exists a mass dependent maximal force and power (6) There exists a maximal energy density and pressure (7) Physical systems must obey the Holographic Principle (8) Holographic bounds can only be saturated by systems with $m>m_p$; systems lying on the ``Compton line'' $l \sim 1/m$ are fundamental objects without substructures (9) We speculate on a new bound from above for the action.\\
In passing, we note that the maximal acceleration is of the order of Milgrom's acceleration $a_0$ for ultra-light particles ($m\sim H_0)$ that could be associated to the Dark Energy fluid. This suggests designing toy-models in which modified gravity in galaxies is driven by the DE field, via the maximal acceleration principle.
\end{abstract}
\maketitle

\section{Three fundamental inequalities}
\label{sec1}
\noindent Usually, one combines the three fundamental constants $G, \hbar$ and $c$  to define Planck's system of units. We believe this should be thought the other way around. In a very simplistic, yet fundamental, conception of our world, all we have is energy that moves through space and evolves in time. It is then natural to assume the existence of three fundamental scales defining these quantities: a (Planckian) mass or energy $m_p$, a length $l_p$, and  duration $t_p$. In this point of view, we may ask why these three fundamental scales have so far manifest themselves through three specific ratios: 
\beq
c=\frac{l_p}{t_p}, \quad G= \frac{l_p^3}{m_p t_p^2}, \quad \hbar= \frac{m_p l_p^2}{t_p},
\label{eq1}
\eeq
and several combinations thereof. This is a non-trivial question. We shall follow the point of view according to which fundamental constants determines limiting properties for physical systems, see for instance~\cite{Duff:2001ba,Uzan:2002vq}. This means that the fundamental constants arise in the various relevant limits of the fundamental theory built upon $(m_p,l_p,t_p)$, when applied to different physical systems characterized by some parameters $(m,l,t)$ (we shall not give any precise definition of a \textit{physical system} in this paper). According to this point of view, the constants $(G,\hbar,c)$ and their combinations emerge from what are probably the most fundamental laws of nature. Their non-trivial expressions as a function of the fundamental scales, Eq.~(\ref{eq1}), thus yield important insights about the laws governing the $(m_p,l_p,t_p)$ world. \\
\\
In this section, we shall discuss three basic inequalities that can be seen as responsible for the emergence of, respectively, $c$, $c^2/G$ and $\hbar/c$. The speed of light is simply a speed that cannot be exceeded, while we argue that $c^2/G \sim 1.3 \times 10^{27}$ kg.m$^{-1}$ (resp. $c^4/G\sim 1.2 \times 10^{44}$ J.m$^{-1}$) is linked to an upper bound for the amount of mass (resp. energy) that is storable in a given region of spacetime. The idea is hardly new, but we analyze it carefully and provide with a new, covariant mass bound conjecture, Eq.~(\ref{l2p}), that is built upon the tools developed by Bousso in~\cite{Bousso:1999xy, Bousso:1999cb,Bousso:2002ju}. The case for $\hbar/c$, although coming from basic quantum laws, is not crystal clear, and we suggest that it might arise as a consequence of another covariant and geometrical inequality relating the enclosed mass and the area bounding the region of interest, see Eq.~(\ref{eqqm}).\\
\\
In Section~\ref{sec2}, we combine these three inequalities to gain some preliminary insights about phenomenological laws of quantum gravity, and discuss several results of the literature. Section~\ref{sece} is a short review of our findings.
\\
\\
Let us first consider the case of the speed of light. Here, fundamental scales combine to define a speed $c=l_p/t_p$ that cannot be exceeded by any physical system in any frame. We then have a first law:
\begin{assumption1}{Existence of a maximal speed.}
Locally, and in any frame, the speed $v$ of physical systems must satisfy:
\beq
v \leq c.
\eeq
\end{assumption1}
\noindent The constant $c$ thus appears as ``fundamental'' because rooted in the kinematical and dynamical structure of the world. For now and then, we write $c=1$ and thus identify $t_p$ and $l_p$. The two following combination then remains: 
\beq
\hbar= m_p l_p, \quad
G= \frac{l_p}{m_p}.
\eeq
Why do they show up in physics, and what can we learn from this? In particular, how do these expressions depend on the number $D$ of spacetime dimensions?\\
\\
There are several evidences that the constant $G$ (in fact $G/c^2$) is also linked to another limiting property for physical systems. Indeed, General Relativity (GR) indicates that event horizons form when one tries to store too much energy in a given region of space-time\footnote{We stay deliberately vague at this point because  the question of the end state of gravitational collapse is a very hard one.}.  Moreover, the horizon will grow in a roughly linear way with the enclosed mass. For instance, in four dimensions and assuming spherical symmetry, a body of mass $m$ and size $l$ must satisfy the bound $ G m/ l c^2 \leq 1/2$. This is the well-known Schwarzschild result, but here interpreted as providing a maximal mass that one may store in a spherical volume in spacetime. As nicely explained in~\cite{Hossenfelder:2012jw}, this understanding of the Schwarzschild solution was already advocated by Bronstein in 1936~\cite{Bronstein:2012zz}. It is then tempting to generalize the idea, and assume that the constant $G=l_p/m_p$ shows up because of some \textit{generalized law that bounds from above the mass\footnote{We prefer to use mass rather than energy since the former is a Lorentz scalar while the latter is not. More comments below Eq.~(\ref{l2p}).} that one may accumulate in a given region of spacetime\footnote{One could also formulate this principle as follow: $1/G$ is the maximal gravitational potential $\phi \sim m/ l c^2$ that can be created by a matter distribution: $G \phi \leq \mathcal{O}(1)$, see e.g.~\cite{Martins:2002fm,Uzan:2002vq}. However, it is not easy to give a covariant meaning to the gravitational potential, and we thus prefer to consider a mass bound.}}. Roughly then, we may require a second law of the following form:
\begin{assumption2}{(First formulation). Existence of a mass bound.}
In $D=4$ spacetimes, any physical system with typical size $l$ has a maximal mass $m$ given by:
\begin{eqnarray}
\frac{G m}{l c^2} \leq \mathcal{O}(1).
\label{law2}
\end{eqnarray}
\end{assumption2}
\noindent Such a law is only indicative, or say, of heuristic interest. Indeed, it has at least two major drawbacks. First,  it is not clear to which ``size'' of the system the quantity $l$ refers to, especially in situations far from spherical symmetry. The physical system could for instance be compact in two spatial directions but very elongated in the third one. One could then define $l$ as the radius of the smallest sphere containing the system. A second, and somewhat related issue, is the fact that the bound is not written in a manifestly covariant way. It actually violates Lorentz symmetry whenever the size $l$ is not given by the areal radius $l=\sqrt{A/16\pi}$ of the cross-sectional area $A$ of a null surface defining an apparent horizon~\cite{Akcay:2007vy}. Using this law outside its domain of validity is then \textit{a priori} misleading, as it may result in spurious Lorentz-violating effects. We shall discuss this in more details in Section~\ref{seclor}. \\
\\
For these reasons, we find that this first formulation for the Law 2 is far too much naive. We believe that a law like Eq.~(\ref{law2}) exists in nature, but we need to write it in a more manifestly covariant way, and moreover without prior assumptions about the symmetries of the system.\\
\\
Quite a similar problem has occurred when people discussed examples and counter-examples to the entropy bound originally proposed by Bekenstein~\cite{Bekenstein:1980jp}. In the late 90's, Bousso provided with a proper formulation of Bekenstein's inequality, resulting in a covariant entropy bound~\cite{Bousso:1999xy, Bousso:1999cb} which is now the formal expression of the holographic principle in the realm of GR~\cite{Bousso:2002ju}. These mathematical tools can be directly used here. On any spacelike surface $B$ of codimension $2$, we shall thus bound the integrated mass (instead of the entropy) on the light-sheets\footnote{We recall that a light-sheet is a null hypersurface with negative expansion, issued from $B$ and which terminates at caustics, where light rays start expanding again.} $L(B)$ issued from $B$ by the area of $B$. We thus postulate the following:
\begin{assumption2p}{(Second formulation).}  Covariant mass bound conjecture.
 In a four-dimensional spacetime which is free of horizons, and given any surface $B$ of codimension 2, the integrated mass on the light-sheets $L[B]$ must satisfy:
\beq
m[L[B]]^2 < \frac{A[B]}{16 \pi G^2},
\label{l2p}
\eeq
\end{assumption2p}
\noindent where we follow Bousso's definitions and notations~\cite{Bousso:2002ju} (and $c=1$ here). The $D \neq 4$ case will be discussed at the end of this section.  Note that quantum effects have been ignored here. Taking them into account will lead to a more refined picture, see Eq.~(\ref{eqqm}) below. \\
\\
\\
Maybe the main issue regarding this bound is that it requires a well-defined notion of (local) mass, which is a difficult question in GR\footnote{As written by Bousso in~\cite{Bousso:1999xy}, ``Local energy is not well-defined in general relativity and (...) it eliminates any hope of obtaining a completely general bound involving mass''.}. It is thus reasonable to expect that an equation like Eq.~(\ref{l2p}) could eventually be understood with the help of the concepts of quasi-local mass or quasi-local energy, see e.g. the review~\cite{Szabados:2004vb}. We also expect that some energy conditions are needed to guarantee the validity of Eq.~(\ref{l2p}). We however leave the ``conjecture''  in this vague formulation, since its discussion would largely beyond the scope of this paper. In passing, we also remark that one may benefit from Bousso's construction to formulate in a more proper way the Hoop conjecture~\cite{Thorne:1972ji, Senovilla:2007dw}. As we shall ignore naked singularities here, we will however not discuss this topic any further.\\
\\
The conjecture Eq.~(\ref{l2p}) is formulated only for spacetimes free of horizons. This is because the inequality is reversed in their presence: this is the Penrose inequality/conjecture, which, loosely speaking, states that in a spacetime with event horizons of total area $A_T$, the mass of the spacetime is at least $16 \pi m^2 \geq A_T$ (for precise formulations and recent accounts of the Penrose conjecture, see e.g.~\cite{Penrose:1973um, Bray:2003ns, Mars:2009cj}; notice also that Penrose inequality is usually not formulated using light-sheets).  Thus we could even strengthen our conjecture and make it a sort of converse to Penrose's one, by writing the following: \textit{a spacetime is free of horizons if and only if for any spacelike surface $B$ in this spacetime, the integrated mass on the light-sheets $L[B]$ satisfies Eq.~(\ref{l2p})}.\\
\\
Schwarzschild's spacetime saturates this bound. The Kerr--Newman (KN) family, which a for long time was believed to be the only possible end state of the gravitational collapse\footnote{Israel-Penrose-Carter no hair conjecture, before several hairy black holes dressed with e.g. gauge fields have been discovered~\cite{Israel:1967wq, Carter:1968rr, Carter:1997im, Bekenstein:1996pn,robinson2009four}.}, obeys~\cite{jacobson1996introductory}:
\beq
A_H= 4\pi\left( 2 m^2 -q^2 + 2m \left(m^2-q^2-J^2/m^2\right)^{1/2}\right),
\eeq
where $q$ and $J$ are respectively the electric charge and the angular momentum of the BH, and $A_H$ the horizon's area. In particular, BHs of the Kerr--Newman family thus satisfy\footnote{For similar inequalities in a more general context, see e.g.~\cite{Dain:2013qia}. Notice also that we limit ourselves to situations where there is indeed an horizon, i.e. we shall in general  ignore naked singularities. As far as the KN family is concerned, it means here that we do not consider overextreme BHs.}:
\beq
4 \pi m^2 \leq  A_H  \leq 16 \pi m^2.
\label{knf}
\eeq
Therefore, in Einstein-Maxwell spacetimes, the less compact BH is the Schwarzschild BH, while the most compact one is the extremal Reissner Nordström BH (with $J=0, q=m, A= 4 \pi m^2$). If we could assume that the final state of any gravitational collapse process results in a BH of the KN family\footnote{For simplicity, we also ignore the cosmological constant here.}, then we would have the following picture. Introducing the areal radius $l=\sqrt{A/4\pi}$, we would have 
\begin{itemize}
\item Any physical system free of horizons lie strictly above the $l=2m$ line.
\item Any physical system with an event horizon have $m\leq l\leq 2m$. 
\item No physical system lie below the $l=m$ line.
\end{itemize}
In particular, one would thus have $ l \geq m$ always (compare with our first formulation of the Law 2, Eq.~(\ref{law2}), where $l$ is now understood as the areal radius of the surface $A$). Combining such a bound with a quantum argument of the form $m l \geq 1$ (see below) would immediately yield, for instance, that any physical system (being a Black Hole or not) must be larger than the Planck length, see Section~\ref{seca}. \\
\\
However, the possibility is still open that there exists hairy, exotic Black Holes, which could be much more compact than $l=m$. Therefore, generalizing the laws we have found so far, we will assume furthermore that there also exists some pure number $\alpha \leq 1$ such that \textit{any possible Black Hole} in nature satisfy the law $\alpha^2 4 \pi m^2 < A_H$, so that one has $l \geq \alpha \, m$ always\footnote{This linear relation is not necessary. We could also simply require that no BH have an area smaller than the Planck one.}.\\
\\
Summarizing our efforts in putting a bound to the energy one may store in spacetime, we shall assume the following picture:  
\begin{equation}
\left\{
      \begin{aligned}
        &16 \pi m[L[B]]^2 < A[B] \quad & \textrm{No black holes}\\
        &4 \pi m^2 \leq  A_H  \leq 16 \pi m^2  \quad & \textrm{ Kerr--Newman family}\\
        & 4 \pi \alpha^2 m^2 \leq  A_H   \leq 16 \pi m^2 & \textrm{Any other BHs}\nonumber
      \end{aligned}
    \right.
\end{equation}
\noindent where only the second line is not an assumption.
\\
\\
Let us now go back to our discussion of the nature of the fundamental constants and consider now the quantum of action $\hbar$. Besides being a conversion factor between energy and frequencies, thus leading to the wave-corpuscule duality, there is indeed a kind of limiting behavior associated to this scale. Quantum fuzziness appears typically when any quantity associated to a physical system, with the dimension of an action, get close to $\hbar$. Stated differently, $\hbar$ is the unit cell of a system's phase space (actually it is $\hbar/2$). This limiting property is expressed through Heisenberg uncertainty principle (position--momenta and/or time--energy). \\
\\
However, and at least to the opinion of the author, the case of $\hbar$ is somewhat less intuitive than the one of $G$ or $c$, and its emergence is rather mysterious. In fact, we will give below an alternative interpretation of $\hbar$ which may lead towards a quite different, more geometrical, understanding of quantum mechanics\footnote{We also note that other radical proposals for new understanding of quantum mechanics are still investigated, see e.g.~\cite{'tHooft:2010zz}}.\\
\\
Let us first be conservative, and assume simply that any quantity $\sigma$ with the dimensions of an action, and associated to a physical system $S$ characterized by $(m,l,t)$, must satisfy $\sigma = m l^2/t \geq \mathcal{O}(\hbar)$. Interpreting $t$ as the duration of observation of the system, we must have $l \leq t$, and we thus get the following law in the $(m,l)$ space only:
\begin{assumption3}{Existence of the quantum of action.}
There exists a positive pure number $\beta$ such that any physical system of size $l$ and mass $m$ satisfies
\beq
 m l  \geq \frac{\beta \hbar}{c}.
\eeq
\end{assumption3}
\noindent The Heisenberg uncertainty principle suggests that $\beta =1/2$. It is worth noting that such a law for $\hbar$ clearly implies that it is independent of the number $D$ of spacetime dimensions.\\
\\
Let us imagine however that we had no knowledge that the law $m l \geq \hbar/c$ derives from the Uncertainty Principle, which itself derives from canonical commutation relations\footnote{Special Relativity, Law 1, has also been used here.}. Then, we would have a heuristic law stating than any system must have a size $l$ greater than its Compton length $\lambda=\hbar/m c$. But it would not be clear in this case to what size the quantity $l$ precisely refers to, especially for systems far from spherical symmetry. Then, by squaring the above law, one would be tempted to see the quantum inequality in a more geometric way:
\beq
m^2 l^2 \geq \frac{\beta^2 \hbar^2}{c^2} \nonumber
\eeq
where $l^2$ has now the dimension of an  area. Again, this can be easily given a manifestly covariant, geometrical form by using Bousso's tools. Let again $B$ be a surface of codimension 2, $A$ its area, $L[B]$ the lightsheets issued from $B$, and $m$ the integrated mass over $L[B]$. We have then the very intriguing result:\\
\\
\textit{The quantum of action in four dimensions can also be interpreted geometrically, as another bound relating the area and the enclosed mass:}
\beq
A> \frac{4 \pi \beta^2 \hbar^2}{m^2 c^2},
\label{eqqm}
\eeq
while its generalization in other dimensions would now lead to a D-dependent $\hbar$ (more on this below). The above quantum inequality should be compared to its gravitational counterpart, namely:
\beq
A > \frac{16 \pi G^2 m^2}{c^4}.
\label{eqgr}
\eeq
The area thus seems to play a special role both in the gravitational and in the quantum world. Also, the correspondence $\hbar/mc \leftrightarrow G m/c^2$ is striking\footnote{This is of course nothing else than the correspondence between the Compton size and the Schwarzschild radius already emphasized e.g. in~\cite{Garay:1994en,Adler:1999bu,Hossenfelder:2012jw} in several contexts, but here written in a manifestly covariant, geometric way.}. Presumably, a more general formula smoothly interpolates between these two regimes around $m\sim m_p$. We note that, taken together, Eqs. (\ref{eqqm}) and (\ref{eqgr}), imply the existence of a minimal area of the same order of magnitude ($\sim l_p^2$) than the one derived in Loop Quantum Gravity~\cite{Ashtekar:1996eg}. The gravitational inequality is rather intuitive since it basically states that storing energy requires space (although the formula is non-extensive, i.e. $m \leq l$ instead of $m \leq l^3$) . The quantum law, at the contrary, is counter-intuitive in this respect.\\
\\
Interestingly, a recent study~\cite{PhysRevLett.110.250502} has derived a formula quite similar to Eq.~(\ref{eqqm}). The authors of~\cite{PhysRevLett.110.250502} show that the product of the average energy of a quantum device by its average area is always bounded by below. The bound is more precise than Eq.~(\ref{eqqm}) in the sense that the authors have computed how it depends on the entropy and the internal number degrees of freedom of the system. Their result is however non-relativistic. This suggests that a more refined bound such as Eq.(\ref{eqqm}) could be obtained, which could be both relativistic, manifestly covariant, and that would also take into account the information carried by the system, in the spirit of~\cite{PhysRevLett.110.250502}.\\
\\
We do not claim, however, that Eq.~(\ref{eqqm}) is the right ``geometrical'' interpretation of the constant $\hbar$ ($\hbar/c$ in fact). The gravitational mass bound is rather intuitive, and arises as a consequence of gravity as being described by spacetime's metric. Similarly, one would need to understand better the physical origin of Eq.~(\ref{eqqm}) in order to give a foundation to such a geometrical reading of $\hbar/c$. Moreover, the generalization to $D \neq 4$ spacetimes is problematic.\\
\\
Let us indeed generalize Eqs. (\ref{l2p}) and (\ref{eqqm}) to $D \neq 4$ cases. The Riemann/Penrose inequality in higher dimensional spacetimes becomes a formula of the form $A_{D-2}^{D-3} \leq G_D m^{D-2}$, where pure numbers have been dismissed here, and $A_{D-2}$ is the area of a codimension 2 surface, and $G_D$ is $D$-dimensional Newton's constant, $G_D= l_p^{D-1}m_p^{-1} t_p^{-2}$. See~\cite{Barrabes:2004rk,bray2009riemannian} for details. Our conjecture would thus generalize to something like
\beq
l \geq m^{\frac{1}{D-3}},
\label{l2d}
\eeq
in terms of the areal radius $l=A_{D-2}^{1/(D-2)}$, and provided $D\neq 3$. As far as the quantum of action is concerned, taking  Eq.~(\ref {eqqm}) seriously and generalizing to $D\neq4$, we would obtain 
\begin{equation}
l > m^{-2/(D-2)},
\end{equation}
while the constant $\hbar$ now explicitly depends on $D$. One finds:
\beq
\hbar_D = \frac{m_p l_p^{D/2}}{t_p},
\eeq
so that only in $D=4$ it has the dimension of an action. If this understanding of $\hbar$ in geometrical terms is any true, then $\hbar_D$ in $D \neq 4$ cannot be anymore the conversion factor between energy and frequencies, nor used to express the canonical commutation relations unless the concept of energy itself becomes D-dependent\footnote{As is transparent when one writes e.g. $[x,p]=i \hbar_D$: in this case the canonical momenta $p$ has not the usual dimension.}. Dynamics would be also considerably impacted. Consider for instance the action for gravity in D dimension, $S_D = 1/G_D \int d^Dx \mathcal{L}_g$, with $G_D=l_p^{D-3}/m_p$, and assume that $[S_D]=[\hbar_D]$. This implies that the gravitational Lagrangian $\mathcal{L}_g$ has dimensions $l^{D/2 -4}$, and thus cannot be the curvature scalar except in $D=4$. All this seems rather unlikely. On the other hand, one might still write standard actions and lagrangians, e.g. $S_D = 1/G_D \int d^Dx \sqrt{-g} R$ where $R$ is the curvature scalar, but in this case these actions would not have the dimension of $\hbar_D$. As a consequence, the Hamiltonian canonical quantization scheme could not be applied anymore. \\
\\
In either cases therefore, a D-dependent $\hbar$ would have a huge impact on extradimensional physics. Again, such a geometrical interpretation of $\hbar$ might simply be incidental in $D=4$, while $\hbar$ has always dimensions $m l$ for any $D$. We shall not explore any further the meaning of Eq.~(\ref{eqqm}) in the present paper, but it deserves more thinking, as it might be a first step towards a more geometrical interpretation of quantum laws (at least in 4D).

\section{Heuristic laws for the quantum gravitational world}
\label{sec2}
\noindent We shall now limit ourselves to a four dimensional spacetime. In the previous section, our attempt to see how the fundamental constants emerge from quantum gravity led us to understand the two ratios $G/c^2$ and $\hbar/c$ in a \textit {structural} way, in the sense that they are associated with non-trivial bounds relating the ``size of a system'' $l$ and its mass $m$. These bounds have been given a manifestly covariant form using Bousso's formalism, in which case $l$ is in fact the areal radius of any codimension 2 surface $B$ and $m$ as the integrated mass on the light-sheets generated by $B$. \\
\\
These inequalities can now be combined to derive some heuristic laws about relativistic quantum and/or gravitational world. We first summarize these results and hypotheses by drawing the physically allowed region in $(l,m)$ space. The use of $l$ instead of $A$ is convenient (although the area has been shown to be much more relevant, physically speaking).

\subsection{On the size and the mass of physical systems}
\label{seca}
\begin{figure}[ht]
\begin{center}
\includegraphics[width=8.5cm]{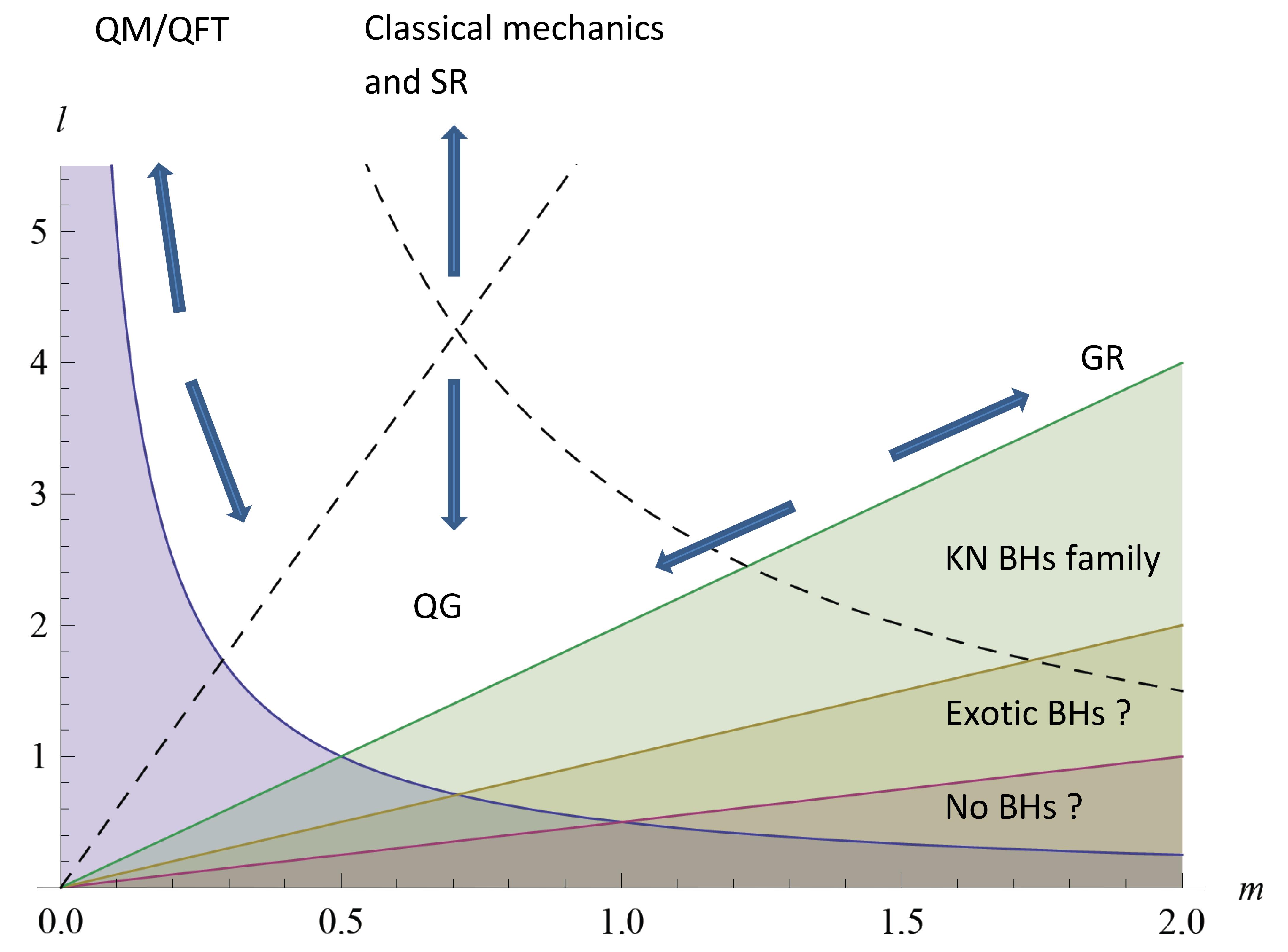}
\end{center}
\caption{Allowed regions for a physical system of size $l$ and mass $m$, in units of $l_p$ and $m_p$. The quantum law forbids any system to enter the $l<1/ (2m)$ region (we have taken $\beta =1/2$ in the graph). According to the Law 2, systems without event horizons live in the region we have left blank. All systems below or on the line $l=2 m$ have an horizon (inspired by the Penrose conjecture). The $l=2m$ line is the Schwarzschild BH, while the KN family lies between $m \leq l \leq 2m$. Below, there might be hairy BHs dressed by some gauge fields, etc. We however assume that there exists $\alpha$ such that any  BH have at least $l \geq \alpha\, m$. Here we have taken $\alpha =1/2$ (bottom line). Under this assumption, it becomes clear from the graph that no systems can be smaller than $\mathcal{O}(l_p)$. Dashed curves corresponds to $l=6m$ and $3/m$ respectively. This enables to visualize roughly the Quantum Gravity (QG) region. Quantum Mechanics (QM) or Quantum Field Theory (QFT) region is reached for $l \gg m$ and $l \sim 1/m$, classical mechanics or Special Relativity (SR) for both $l \gg m$ and $l \gg 1/m$, while GR corresponds $l \gg 1/m$ and $l/m = cst$.}
\label{fig1}
\end{figure}

\noindent The bounds Eqs. (\ref{l2p}, \ref{knf}, \ref{eqqm}) are shown in Fig. (\ref{fig1}). Taken together, they delimit a physically allowed region in the $(m,l)$ space. The physical region can be further split in subregions corresponding to physical objects well described by the relevant limit of the full theory.  \\
\\
By combining the Schwarzschild inequality together with the quantum one, many papers of the literature have come to the conclusion that physical systems cannot become smaller than the Planck size (see references in the next subsection). As made clear by Fig. (\ref{fig1}), we basically agree with this conclusion, although we wish to stress the following points.\\
\\
If by the wording  \textit{physical system}, we understand any system that has not turned into a Black Hole, then the Fig. (\ref{fig1}) makes transparent that any physical system must indeed be at least as large as the Planck size (up to some pure number of order one). This is obtained combining the covariant mass bound conjecture (Law 2) together with the quantum inequality (Law 3). \\
\\
However, there is no reason to exclude Black Holes from the set of physical systems. Then, a more careful analysis is required since BHs are the only physical systems allowed to cross the $l=2m$ line. It is also easily seen from Fig. (\ref{fig1}) that no BH of the Kerr--Newman family can become smaller than the Planck length (again, up to some different number of order 1). However, we know no way of proving that there cannot exist an exotic, hairy Black Hole with a large mass, but very small size (a classical object, in the sense that $l\gg1/m$, but $m \gg 1$ and $l \ll1$). Of course, this seems rather odd, and this is why we have also assumed that there exists $\alpha \leq 1$ such that $l \geq \alpha \,m$ for any physical systems. Using then $l \geq \alpha \, m$ and $m l \geq \beta$ obviously results in a minimal size for physical systems $l \geq \sqrt{\alpha \beta}$, so that, under this further assumption, we agree with the literature. However one should keep in mind that the existence of a mostly classical, heavy, and subplanckian BH would invalidate this belief.\\
\\
 In the following, and for the sake of clarity, we shall now set $\alpha = \beta =1$, and use Law 2 and 3 in the simple form $m l >1$ and $m< l$. We also use strict inequalities for a better reading, although it should be understood in fact as ``inequalities in order of magnitude'':  $\lesssim$ and $\gtrsim$.\\
\\
We also note that  Fig. (\ref{fig1}) shows where is the Quantum Gravity (QG) regime in the $(l,m)$ diagram: it is a rather narrow space that can be reach, it seems, in three different ways. A first one is to consider a quantum mechanical/QFT system (always staying close to the Compton line $l \sim 1/m$), and reduces its size down to the Planck length (i.e., increasing its mass at the same time). A second well-known way is to follow an evaporating Black Hole until it reaches a mass of order the Planck mass. The third way is to look at a fixed-mass system, and study it in a smaller and smaller box. This last way to quantum gravity has been already discussed in~\cite{Rovelli:2008cj}. More precisely, the author of ~\cite{Rovelli:2008cj} argues that the region that we have labeled ''Classical mechanics and SR'' in Fig. (\ref{fig1}) may be further split, with the region given by $m\sim m_p$ but large $l$, as being the region covered by DSR (and presumably also GUP) theories. This would describe classical systems in the sense that $l\gg 1/m$ and $m \ll l$, but $m\sim m_p$, i.e. systems controlled by physical effects, if any, associated to the constant ratio $\hbar/G$ with $\hbar \to 0$ and $G\to 0$. 

\subsection{Minimal Length Scenarios and the question of  Lorentz symmetry}
\label{seclor}
\noindent Combining both Laws 2 and 3, we have derived the existence of a \textit{minimal area} which is mass dependent, i.e. a minimal ``size'' for physical systems. As we have said before, many papers of the literature have already come to this conclusion, basically using these two ingredients in several quite convincing thought experiments. More precisely, what has been derived roughly split as follow (this is by means an exhaustive list).\\
\\
First, and in a fashion very similar to the Fig. \ref{fig1}, one shows that a bound system cannot be compressed below the Planck length. This is trivially obtained as consequence of the fact that the energy within some (spherical) box of size $l$ is $E >1/l$ (by virtue of the uncertainty principle), and using next the Schwarzschild bound, see e.g. \cite{Mead:1964zz,Padmanabhan:1987au}. \\
\\
Second, one may derive the existence of a minimal resolution, of order the Planck scale, to position measurements in space and time. This can be derived using the Law 2 and 3, i.e. the fact that the formation of micro Black Holes and event horizon ultimately limits our ability to probe spacetime below the Planck scale, see e.g. \cite{Doplicher:1994tu, Scardigli:1999jh, Hossenfelder:2012jw, Calmet:2005mh}. Here the topic is however somewhat controversial. Most studies indeed implicitly assume spherical symmetry, and hence consider the simultaneous measurement of the three coordinate positions. In situations far from symmetry, it has been argued that only specific combinations of coordinate uncertainty can be derived \cite{Doplicher:1994tu}; in particular, an unlimited accuracy on the measurement of one coordinate does not seem excluded by first principles, see the debate in \cite{Hossenfelder:2012kk,Doplicher:2012en} and references therein. We believe our covariant formulation of the Law 2 may help further in clarifying this issue, since it shall also apply in highly non symmetric situations.\\
\\
Finally, one can also derive a minimal resolution to spacetime measurements by using only the  ``gravitational push'' exerted by the probing particle on the system to be probed\footnote{For a very clear demonstration of that point, see \cite{Adler:1999bu}. The authors consider a tube of light of finite extent hitting a test-particle. The particle gets a gravitational kick when hit by the tube of light, and an opposite kick when it leaves. This results in a net displacement that must be added to the standard uncertainty.}, e.g. in scattering thought experiments like the Heisenberg microscope \cite{Mead:1964zz,Padmanabhan:1987au,Garay:1994en,Adler:1999bu} (see also \cite{Maggiore:1993rv} for a different reasoning with similar results). Quite interestingly with respect to the above controversy, these thought experiments lead to a minimal resolution without relying at all on horizon formation. One of the main lesson of these studies is the fact that --and contrary to a widespread belief originating from the early days of quantum mechanics-- a wave of wavelength $\lambda$ cannot be used to probe spactime with a resolution $\Delta x \sim \lambda$, but only with a resolution $\Delta x \sim \lambda + l_p^2/\lambda$. This is because of this gravitational push which cannot be ignored anymore near or above the Planck scale. Thus, even if subplanckian wavelength are authorized (more on this in the Section \ref{secb}), such a wave cannot be used to probe spacetime with an accuracy better than the Planck scale.  The other, more technical, outcome is a Generalized Uncertainty Principle (GUP) of the form 
$$
\Delta x \Delta p > f(\Delta p ^2),
$$
where the standard GUP is $f(x) = 1+ \beta x$ for some positive number $\beta$. The GUP can be derived from modified canonical commutation relations, see e.g. \cite{Kempf:1994su}.\\
\\
These ideas and models constitute, generally speaking, the Minimal Length Scenarios, see \cite{Hossenfelder:2012jw} for a recent review. The modified commutation relations break Lorentz symmetry. In parallel, and more generally, it has also been claimed that new physics associated to the Planck scale require the Lorentz symmetry to be broken or at least deformed, in such a way that the Planck length (or the Planck energy), becomes another fundamental invariant. It led to the study of the so-called Doubly or Deformed Special Relativity (DSR) theories, which deforms the symmetry group to include both an invariant speed and an invariant length scale~\cite{AmelinoCamelia:2000mn,AmelinoCamelia:2000ge}, or an invariant energy scale~\cite{Magueijo:2001cr}.\\
\\
However, we do not \textit{a priori} agree with the inevitability of Lorentz symmetry breaking near the Planck scale.  Indeed, and as anticipated in the first section, it is rather natural to end up with non-covariant result when one uses the Law 2, and promoting  the quantity $l$ in the equation Eq.~(\ref{law2}) to a generic ``size'' of the system, or as being the typical extent of the interaction region in a scattering process. This comment especially applies to the reasoning based on the formation of horizons. Thus, we do not claim that Lorentz symmetry remains intact near the Planck scale, but we believe that a more careful treatment of these thought experiments is required, using a manifestly covariant criteria for horizon formation. Moreover, it has been argued in \cite{Rovelli:2002vp} that quantum operators associated with spatial measurements might well admit a minimal proper value without threatening Lorentz symmetry, because such an operator shall not commute, in general, with its boosted counterpart\footnote{In \cite{Rovelli:2002vp}, only area operators are considered.}. The fate of Lorentz symmetry near the Planck scale is thus rather unclear, at least to the author's opinion. In the next section, we even show that the basic inequalities discussed in the first section do not imply that a wave must have a minimal wavelength, quite in contradiction with the whole spirit of DSR. 
  
\subsection{Transplanckian frequencies are allowed}
\label{secb}
\noindent Although physical systems have a size which is bounded by below, we now provide with a simple argument showing that the three inequalities discussed in the previous section do not  imply the existence of a minimal Planckian \textit{wavelength}. \\
\\ Let us indeed consider the case of a monochromatic wave/quanta of wavelength $\lambda$, in a cubic box of linear dimension $l$. The quanta needs to fit within the box, so that there exists an integer $n \geq 1$ such that $\lambda = l/n$. The quanta has an energy $m=E=n/l$. The Law 3 is satisfied since $n \geq 1$, while Eq.~(\ref{law2}) reads:
\beq
\frac{m}{l}=\frac{n}{l^2} < 1, \nonumber
\eeq
from which we derive $l^2 > n \geq 1$. This shows that the physical system (the box containing the quanta) is indeed larger than the Planck length, as expected from the previous discussion. However, using that the quanta wavelength reads $\lambda=l/n$, we also get:
\beq
\frac{l_p}{l} < \frac{\lambda}{l_p} < \frac{l}{l_p},
\label{eqlambda}
\eeq
or, in terms of the frequency:
\beq
\frac{l_p}{l} < \frac{w}{w_p} < \frac{l}{l_p}.
\label{eqw}
\eeq
The non-trivial part of these inequalities, coming from the Law 2, are $\lambda > 1/l$ or equivalently $w < l$. Thus the lower bound for the quanta's wavelength is $l_p^2/l$, which can be much smaller than $l_p$ when box is large enough. As far as Laws 2 and 3 are concerned, there is no dramatic physics (such as a collapse of the wave into a ``black wave'', so to say) when the quanta's wavelength become subplanckian. It is only when  $\lambda$ approaches $l_p^2/l$ that the box which contain the quanta saturates the mass bound Eq.~(\ref{law2}), and maybe collapse into a Black Hole\footnote{In this case, it must be true in any frame.} (here we only use the Law 2 in its naive form, Eq.~(\ref{law2})). Such a result is actually confirmed by the exact electromagnetic plane waves solutions to Einstein-Maxwell system of equations, the pp-waves spacetimes. These spacetimes (in our case, the one describing an electromagnetic wave filling all the space, i.e. with $l \to \infty$) exhibit no singularity, whatever the wavelength \cite{misner1973gravitation}. Again here, there are no ``Black Waves''.\\
\\
On physical grounds, these transplanckian frequencies are allowed because what matters for Einstein equations is not the energy of quanta, but the energy density in the box. Since the quanta is essentially delocalized in the box, the average energy density can be low whereas the energy of the quanta is large. Explicitly, the energy density in the box reads $\rho= E/l^3 = n/l^4$. Quantum Field Theory is expected to be an appropriate description for the physics inside the box provided its self-gravity is small: $E \ll l$, or, equivalently, provided the curvature $L$ generated by the box is much larger than the box itself $l \ll L$. These conditions are equivalent, and one finds again $1/l \ll \lambda < l$, which roughly defines the domain of validity of QFT in a box for one particle  (compare to Eq.~(\ref{eqlambda})). \\
\\
 In passing, we remark that the UV and IR cut-off must thus be related to each other. This is a well-known consequence of the Holographic Principle which will be briefly discussed below. We note however that our bound defining the domain validity of QFT does not agree with the one discussed in, e.g.~\cite{Cohen:1998zx}. This is because we have limited ourselves to only one particle. If we consider instead $N$ particles \textit{with the same energy} $w$ in our box, then their maximal frequency now reads $w < l/N$. In the reasoning of effective field theory where one looks at a theory up to some UV cut-off $w_{\textrm{max}}=\Lambda$, the maximal number of degrees of freedom is $N=(l \Lambda)^3$. Then the Law 2, which reads $w_{\textrm{max}}<l/N$, gives $l^2 \Lambda^4 <1$. One concludes that QFT is a valid description if the box is not too large, i.e. for a maximal IR cut-off $l$ related to the UV one by $l \lesssim 1/\Lambda^2$, as was indeed found in~\cite{Cohen:1998zx}. However, this last result must be understood in the point of view of effective field theory where highly energetic processes have been integrated out. Physically speaking, whatever the cut-off $\Lambda$ is, there can always exist very energetic quanta $w \sim l$ that live above that cut-off. The maximal energy a quanta may have in a box seems to depend in fact on both depends on the box's size, the number of particles, and their equation of state.\\
\\
One might find paradoxical that physical systems cannot be smaller than the Planck scale while subplanckian wavelength are allowed. We think it is no paradox, for it would be misleading to think of a wave of wavelength $\lambda$ as a physical system of size $\lambda$. Instead, such a wave is completely delocalized, so that what matters as a physical system is not the wave itself, but the box that contains it. The box cannot be smaller than the Planck scale, while the quanta it contains can have a transplanckian frequency. Moreover, it does not conflict the existence of a minimal resolution to space-time measurements, since a subplanckian wavelength cannot be used to probe distances smaller than the Planck length, as recalled in the previous section. Conversely, a minimal resolution such as the one derived in the GUP framework does not necessarily imply a minimal wavelength\footnote{Assume indeed a one dimensional commutation relation of the form $[x,p]=i \hbar f(p)$ for some function $f$. Define a fictitious momentum $\pi=g(p)$ where $g^{-1}(p)= \int dp/f(p)$, so that $[x,\pi]=i \hbar$. Then the standard wave-corpuscule duality applies to $\pi$, and reads $\lambda=h /\pi$. The modified De Broglie relation is thus $\lambda= h/g^{-1}(p)$. Then one shows that a minimal $\Delta x$ is not equivalent to a minimal wavelength. For instance, in the model $[x,p]=i \hbar(1+ \sqrt{G/\hbar}\sqrt{p^2})$, there is a minimal $\Delta x$ but the wavelength is not bounded by below, as one can check.}.\\
\\
A final comment regards the Lorentz-(non)-covariance of Eqs. (\ref{eqlambda}) or (\ref{eqw}). As we have anticipated in Section~\ref{sec1}, the use of the Law 2 under its naive form $m/l \leq 1$  leads, without surprise, to non-covariant formulas. We have commented this at length in the first section, and put forward a covariant expression for the mass bound. We thus believe that equations like Eqs. (\ref{eqlambda}) or (\ref{eqw}) should not be understood in their strict sense, but only as indicative. They point towards a more thorough discussion of the domain of validity of QFT. It is striking that we find effects that depends on the box's size itself. Such a non-locality suggests a radical revision of the nature of degrees of freedom in QFT plus gravity, pretty much as the holographic principle does~\cite{'tHooft:1993gx, Susskind:1994vu}, see also Section~\ref{secholo} below.

\subsection{Transplanckian momenta}
\noindent The simple relation $p=\hbar k$ together with results of the previous section show that a quanta in a box of size $l$ must also obey
\beq
\frac{l_p}{l} < \frac{p}{m_p} < \frac{l}{l_p}
\label{eqsp}
\eeq
It cannot go to zero because of the Heisenberg principle, but can be transplanckian in a large enough box. Similar remarks about Lorentz covariance apply here.

\subsection{Existence of a maximal acceleration}
\begin{figure}[ht]
\begin{center}
\includegraphics[width=4cm]{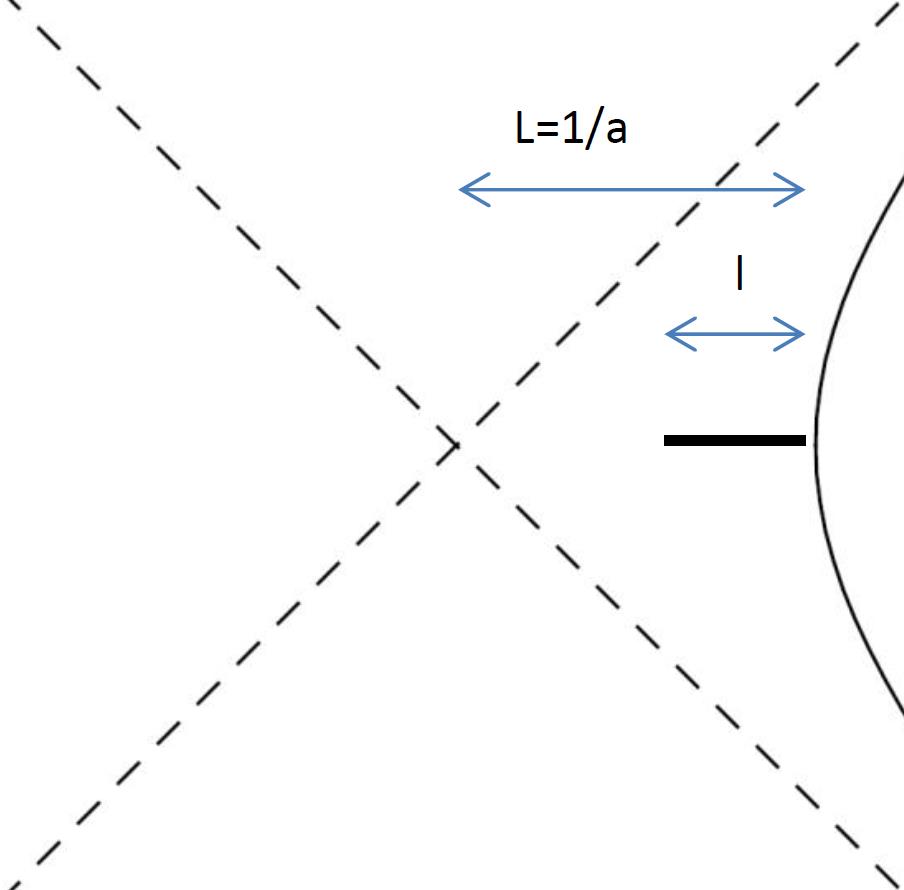} \hfill
\includegraphics[width=4cm]{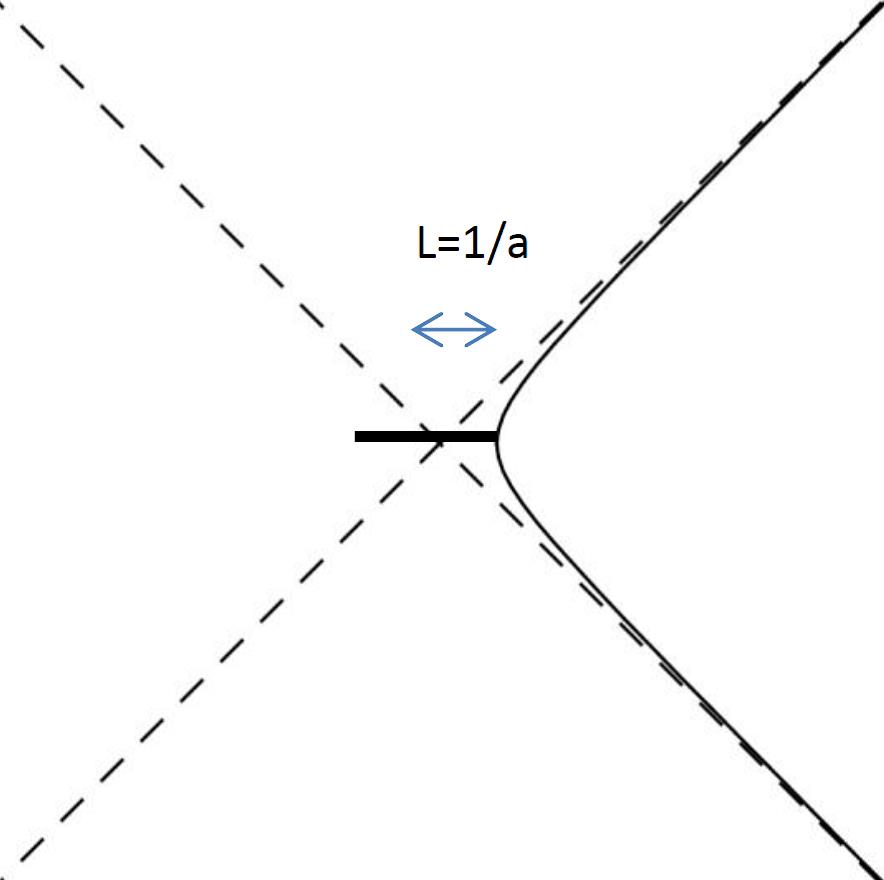}
\end{center}
\caption{Minkowski diagram of the same system $S$, of size $l$ (thick segment), undergoing different constant proper acceleration along some direction $x$. Left: the proper acceleration $a=1/L$ is such that $L>l$. Right: for a larger acceleration, $L< l$, and some parts of the system lie beyond the Rindler horizons.}
\label{fig3}
\end{figure}

\noindent The idea that a maximal acceleration should arise in quantum gravity is not new. Fundamental constants yield a Planckian acceleration $a_p = (c^7/\hbar G)^{1/2} \sim 5\times 10^{51} m.s^{-2}$. Consider again a physical system $S$ of size $l$ and mass $m$. There is a natural acceleration scale associated to this system, namely $a=c^2/l$. Since $l>1$, this acceleration scale is indeed bounded from above $a < 1$. However suggestive, this can hardly be consider as a result, because the system under consideration can well be not accelerating at all.\\
\\
Rather, let us suppose that the system $S$ accelerates constantly at proper acceleration $a$. Let $L$ be given by $L=1/a$. In a Minkowski diagram, two different cases arise, depending on whether $L>l$ or $L<l$. The latter case seems problematic since, here, parts of the system $S$ actually lie above its past and future Rindler horizons, see Fig. \ref{fig3}. Therefore, by accelerating too much a \textit{finite-size} object, internal parts of the system get causally disconnected from each other. The object is thus probably torn apart. What happens precisely is not clear, but it is rather natural to assume that it cannot happen.
Similar arguments can be found for the maximal acceleration of a string~\cite{gasperini1991kinematic,frolov1991instability,gasperini1992causal}. We should thus require $L>l$, i.e. $a<1/l$. 
We thus obtain that no physical systems can accelerate at more than the Planckian acceleration
$$
a< \frac{1}{l} < 1,
$$
where, in the following, we set $m_p=l_p=t_p=1$. In fact, by making use of the physically allowed region discussed in Section~\ref{seca},  we can be much more precise. Indeed we have:
\begin{equation}
  \left\{
      \begin{aligned}
  a&<\frac{1}{l} < m<1 \quad&\textrm{if}& \quad m<1\\
        a&<\frac{1}{l}<\frac{1}{m}<1 \quad&\textrm{if}& \quad m>1
      \end{aligned}
    \right.
\label{eqsA}
\end{equation}
Here it is interesting to insert back the dimensionfull constants. We find two different bounds:
\begin{equation}
  \left\{
      \begin{aligned}
          a&<\frac{m c^3}{\hbar} \quad&\textrm{if}& \quad m<1\\
          a&<\frac{c^4}{G m} \quad&\textrm{if}& \quad m>1
      \end{aligned}
    \right.
\label{eqsAh}
\end{equation}
\begin{figure}[ht]
\begin{center}
\includegraphics[width=8.5cm]{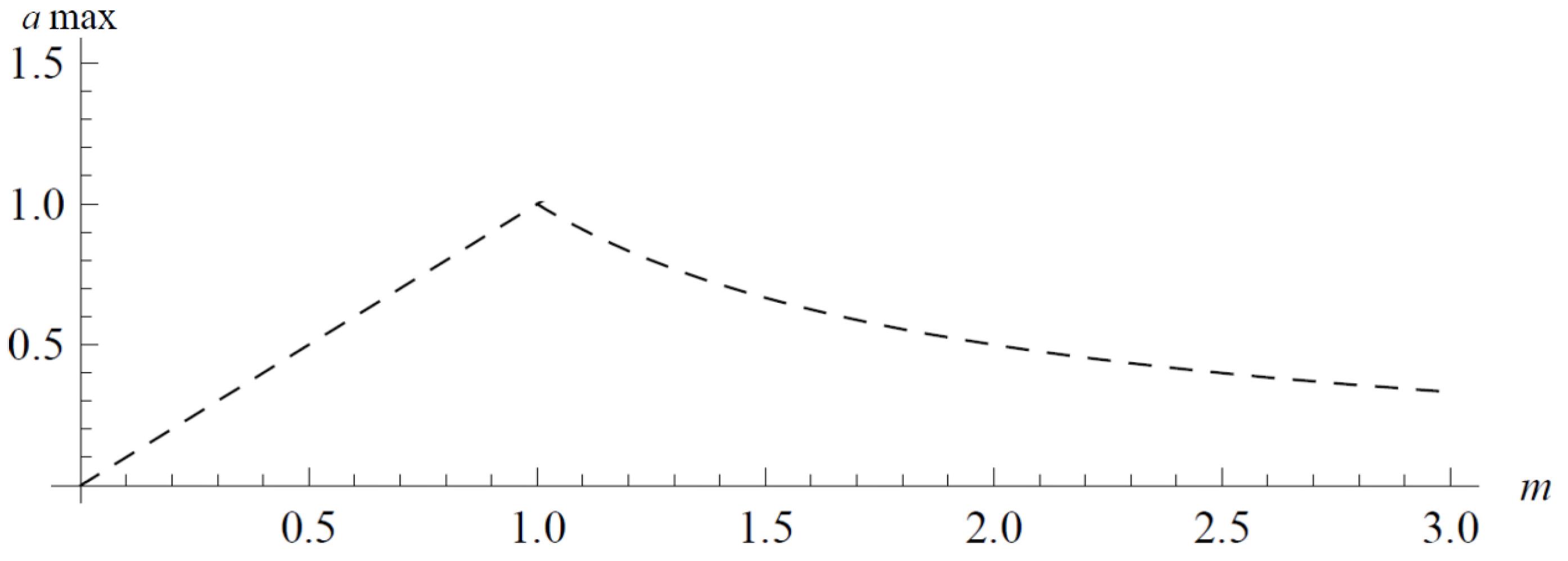}
\end{center}
\caption{A physical system $(m,l)$ cannot have an acceleration greater than $m$ if $m<1$, or $1/m$ if $m>1$. In particular, it cannot have more than a Planckian acceleration $a<1$, which can be reached only by systems with a  Planckian mass.}
\label{fig4}
\end{figure}\\
\\
The first bound is known as Caianiello's maximal acceleration~\cite{Caianiello:1981jq,Caianiello:1982zz,Feoli:2002dva}, which, very interestingly, does not depend on $G$. It has been first derived by Caianiello more than thirty years ago, using quite different, yet very nice arguments. In an attempt to geometrize of the quantum phase space, one assumes a generalized metric of the following form
\beq
d\tilde s^2=-dt^2+dx^2 + \frac{\hbar^2}{m^4 c^6}\left(-dE^2 + dp^2\right)
\eeq
in the eight dimensional space $(x^{\mu}, p_{\mu})$. This yields, in particular, a mass \textit{and} velocity dependent maximal acceleration. The bound is maximal for a body at rest, in which case one has: $a_{\textrm{rest}} < \mathcal{O}(mc^3/\hbar)$. It has also been derived by using the uncertainty principle~\cite{Wood:1989qu,Papini:2003xw}. Many interesting papers on the topic have then followed. In particular, one notes that the metric and the geometry now becomes observer-dependent, leading for instance to mass-dependent corrections to the Rindler metric~\cite{Caianiello:1989wm}. \\
\\
The theory can alternatively be written by assuming the following form for the invariant proper time $d\tau$:
\begin{equation}
-d\tau^2=dx_{\mu}dx^{\mu} + \frac{\hbar^2}{m^2 c^2} d\dot{x}_{\mu}d\dot{x}^{\mu}, ,
\label{eqcaia}
\end{equation}
where the worldline $x^{\mu}$ is parameterized by $s$, with $ds^2=dx_{\mu}dx^{\mu}$, and noting that $ds^2$ is not an invariant anymore~\cite{Caianiello:1989wm}. Such a theory thus violates the clock hypothesis, according to which the proper time depends only on the instantaneous velocity of a material body, but not on higher derivatives~\cite{PhysRevA.47.3611}. The theory of Kawaguchi spaces~\cite{kawaguchi1937theory} provides with adapted tools for studying curves whose arc element depend on higher order derivatives~\cite{Nesterenko:1994tb,Nesterenko:1998jt} (but see also ~\cite{Schuller:2002fn} for a different mathematical formulation).\\
\\
An immediate criticism of such theories is that it shall lead to higher order theories, which are known to be unstable according to Ostrogradski's theorem  (for a short account of the theorem, see e.g.~\cite{Woodard:2006nt}). However, it is likely that not only the acceleration is bounded from above, but also the jerk, and all higher derivatives of the position. Indeed, the argument presented in Fig.~\ref{fig3} should also apply to trajectories with constant n-th derivative of the position. On dimensional grounds, one shall find $a < 1/l$, $\dot{a} < 1/l^2$, and so on\footnote{The exact calculation is technically difficult, however, because solving for e.g. constant proper jerk in special relativity leads to a cumbersome differential equation, while the situation becomes even harder for higher orders.}, and therefore, for $m<1$, one shall have the bounds $a<m$, $\dot{a}<m^2$, $\ddot{a}<m^3$, etc., up to some pure numbers. This shall be in turn modelled via a generalization of Eq.~(\ref{eqcaia}), of the form:
\begin{equation}
-d\tau^2=dx_{\mu}dx^{\mu} + \frac{\hbar^2}{m^2 c^2} d\dot{x}_{\mu}d\dot{x}^{\mu}
+  \zeta \frac{\hbar^4}{m^4 c^4} d\ddot{x}_{\mu}d\ddot{x}^{\mu}+ \ldots
\end{equation}
eventually leading to a \textit{non-local} inertial physics. This could be a cure to the aforedmentioned problem, since infinite order theories can be stable, although their truncated, finite-order counterparts, cannot, see~\cite{Simon:1990ic} (but see also~\cite{Woodard:2000bt, Woodard:2002bx}).\\
\\
Let us also stress that a maximal acceleration of order $a_p$ has been recently derived in Loop Quantum Gravity~\cite{Rovelli:2013osa}. The inverse acceleration (squared) is shown to be related to the measure of an area, whose spectrum is discrete in LQG. Could the LQG somehow provide with bounds similar to Eqs. (\ref{eqqm}) or (\ref{eqgr}) for the expectation values of areas, that this result would agree with ours, and in particular with Caianiello's maximal acceleration.\\
\\
These previous work do not separate between the $m<1$ and $m>1$ cases. As far as we are aware therefore, our second bound, $a<c^4/Gm$ is new. The two bounds Eq.~(\ref{eqsAh}) are much more stringent than $a<a_p$ alone, especially in the very small mass or very large mass limits, see Fig.~\ref{fig4}. For an electron, we find $a_{max} \sim 10^{29} m.s^{-2}$. For the lightest known particles, namely neutrinos with a mass of, say $1$ eV, we find $a_{max} \sim 5 \times 10^{23} m.s^{-2}$, which is still very high. Several phenomenological implications have been discussed e.g. in~\cite{Caianiello:1989pu, Feoli:1997zn,Lambiase:1998tn,Feoli:1999cn, Bozza:2000en, Papini:2002vg}.\\
 \\
Interestingly however, we note that this maximal acceleration scenario may be of some relevance in the context of Dark Energy/Dark Matter issues. Indeed, if the Universe is filled with a Dark Energy scalar field of very light mass, given roughly by the Hubble constant today ($m_{DE}\sim H_0$), then the maximal acceleration for such an ultra-light particle would read $a \sim  H_0 t_p a_p \sim 7 \times 10^{-10} m.s^{-2}$. This turns out to be very close to Milgrom's acceleration $a_0$ which have been shown to play a very special role at astrophysical scales, in particular galactic ones, see e.g.~\cite{Milgrom:1983ca, Bekenstein:1984tv,Milgrom:1992hr, Sanders:2002pf, Bekenstein:2004ne, Bruneton:2007si, Milgrom:2008rv,  Skordis:2005xk}, and~\cite{Famaey:2011kh} for a comprehensive and recent review. \\
\\
The fact that Milgrom's acceleration is of the same order of magnitude than the Hubble constant today ($a_0 \sim H_0$) has been noticed a long time ago. What is new here is that we provide with a natural bound for the acceleration of particles of mass $m<1$, given by $a<m$. When one approaches this bound, modified dynamics must take over, pretty much like Special Relativity modifies Newton's second law to ensure that $v<c$ always hold. As speculative as it may be, we are thus  led to think that  modified gravity behavior à-la-MOND might well be linked to the existence of ultra-light particles of mass $m\sim H_0$ that reach their maximal acceleration (around $a_0$) in a galactic context. What would happen in the Solar System for instance where acceleration is much larger is however unclear\footnote{Not mentioning moreover that acceleration due to gravity is an ill-defined concept in general.}. It is thus preferable to investigate further the idea in the context of field theories. One will get in \cite{Nesterenko:1998jt} a glance at how this might be done. Note that such models would not solve the coincidence problem, which in this context would read: why the scalar partner to gravity has a mass of the order of the Hubble constant today? In any case, we believe that this idea deserves more attention in the light of both Dark Matter/ Dark Energy problems.\\
\\
On the very large mass limit, we find a maximal acceleration given by
\beq
a_{\textrm{max}} \sim \left(\frac{M_\odot}{M}\right) 6 \times 10^{13} \,\,m.s^{-2},
\eeq
where $M_\odot$ is the Sun's mass. Therefore the maximal acceleration for clusters is typically around, or slightly below $1 \,m.s^{-2}$, which is huge in an astrophysical context. The hypothesis is thus hardly testable with large scale structures.

\subsection{Maximal force and maximal power}
\noindent The existence of a maximal force $f_{max} \sim c^4/G$ and power $P_{max} \sim c^5/G$ has been advocated several times in the past, but not often, see e.g.~\cite{Gibbons:2002iv}. Again here, we will provide with more refined bounds, using $l>1/m$ in the $m<1$ case, but $m<l$ if $m>1$.\\
\\
Simply multiplying Eqs.~(\ref{eqsA}) by $m$, we get, using $f=m a$,
\begin{equation}
  \left\{
      \begin{aligned}
      f&<\frac{m^2 c^3}{\hbar} \quad&\textrm{if}& \quad m<1 \\
       f&< \frac{c^4}{G} \quad&\textrm{if}& \quad m>1 
      \end{aligned}
    \right.
\label{eqsf}
\end{equation}
Notice that physical systems thus obey \textit{both} a maximal acceleration and a a maximal force principle. This is not paradoxical because the maximal force $f_{\textrm{max}} = 1$ only applies to systems with $m>1$, while the maximal force in the low mass regime $<1$ is much smaller, with $f_{\textrm{max}}=m^2$, so that $a_{\textrm{max}}=m$ in this case.\\
\\
The fact that the maximal force or power do not depend on $\hbar$ (if $m>1)$ is intriguing. Does it mean that GR already include these principles? If so, can we actually prove them? That would require, to start with, a proper definition of force and power in GR. We believe this deserves more attention.

\subsection{Maximal energy density and maximal pressure}
\noindent A system $S$ has an average energy density given by
\beq
\rho_S = \frac{m}{l^3}
\eeq
Using $m< l$ and $l>1$, we thus have $\rho_S = m/l^3<1/l^2<1$, showing that no system can have an energy density greater than the Planckian density. More stringent bounds can again be found by separating between $m<1$ and $m>1$ systems. One gets
\begin{equation}
  \left\{
      \begin{aligned}
        \rho_S&<\frac{1}{l^2}<\frac{1}{m^2}<1 \quad&\textrm{if}& \quad m>1\\
        \rho_S&<\frac{1}{l^2}< m^2 <1 \quad&\textrm{if}& \quad m<1
      \end{aligned}
    \right.
\label{eqsrho}
\end{equation}
Again here, the bound $\rho_S <1/l^2$ shall be understood in the sense of Bousso's tools, $\rho_S < 1/A$, and up to some pure numbers. Because the pressure is also an energy density, similar formulas should hold for the pressure. In passing, such a  bound on the pressure $P_S < 1/l^2$ guarantees that the force it exerts at the boundary of the object $S$ is less than the maximal allowed force $f = P_S l^2 <1$, in agreement with the maximal force scenario.  

\subsection{Entropy and Holography}
\label{secholo}
\noindent If a system $S$ of size $l$ and mass $m$ is made of $N$ degrees of freedom with an energy $\epsilon = w$ each, then  $N \sim m/w$, and, according to Eq.~(\ref{eqw}), this implies the following inequalities for the number of internal degrees of freedom:
\beq
\frac{m}{l} < N < m l 
\label{eqN}
\eeq
The left hand side does not bring much information, as we know that $m/l <1$, meaning thus that $N > \mathcal{O}(1)$ as expected. The right hand side is already quite similar to Bekenstein's entropy bound~\cite{Bekenstein:1980jp}, $N \sim S \leq 2 \pi E R$, where here $E=m$ and $R=l$ would be the radius of the smallest sphere containing the system. Using next the law $m<l$, we also get
\beq
\mathcal{O}(1)<N< l^2
\eeq
which is the holographic principle in its naive form (as well-known, its covariant expression is $S \leq A/4$, see~\cite{Bousso:1999xy, Bousso:1999cb,Bousso:2002ju}). We thus don't derive anything new here. However, we wish to point out that, in the $m \ll 1$ region, one also has $l\geq1/m \gg m$, and thus 
\beq
1<N\ll l^2 \quad \textrm{if} \quad m \ll 1
\eeq
We conclude that the holographic bound can only be saturated for heavy systems with $m>1$. Finally, for systems which lie on the ``Compton line'' $l=\mathcal{O}(1/m)$, Eq.~(\ref{eqN}) even becomes
\beq
\mathcal{O}(1) < N < \mathcal{O}(1),
\eeq
thereby showing that particles lying on the Compton line are most probably fundamental objects, in the sense that they don't have any internal structure ($N=1$), although they might have a large size with respect to the Planck length.

\subsection{Is the action bounded from above?}
\noindent In deriving all the previous results, we have made heavy use of the quantum inequality $\sigma = m l \geq 1$, requiring that physical quantities with the dimensions of an action should be larger than the quantum of action $\hbar = m_p l_p$. Using also the RG-like inequality $m \leq l$, we can actually bound from above this quantity $\sigma$. This seems to have passed unnoticed so far in the literature. We would then have, for $\sigma = m l$
\beq
\mathcal{O}(1) \leq \sigma \leq \mathcal{O}\left(l^2\right).
\eeq
Hence the action, or say, physical quantities with the dimension of an action, should not be larger than the ``area'' of the system under consideration. 
Let us be more precise here, and consider a system $S$ of mass-energy $m$, size $l$, that evolves during $t$. Then $\sigma = m l^2/t$ shall obey, using $m<l$, 
\beq
\mathcal{O}(1) \leq \sigma \leq \mathcal{O}\left(\frac{l^3}{t}\right)
\label{eqaction}
\eeq
This is clearly quite a strange inequality, for the classical action can be shifted by a constant without impacting the physics. In any event, let us test the bound against some known physical systems. For instance, in flat, matter-dominated cosmologies, the classical action (GR plus matter), integrated over some volume $l^3$ during some duration $t$ reads $S \propto \rho t l^3$ where $\rho$ is the comoving density. Then the previous bound implies $\rho < 1/t^2$. Since the density scales roughly as the Hubble parameter, this bound is in in fact $t < \mathcal{O}(1/H)$ at every epoch, although pure number quite different from $1$ could show up. The meaning of the above bound is thus not very clear; in this case, it seems to provide the range over which time integration in the action can be performed safely. We have also investigated the case of a static, constant density star in Newtonian gravity. In this case, the action, integrated until the boundary of the star, and over some duration $t$ reads $S= t m^2/ (4 l)$, where $m$ is the mass of the star, and $l$ its size. In this case the bound Eq.~(\ref{eqaction}) reads $t< l^2/m$ whose significance is not clear either. 

\section{Summary and outlook}
\label{sece}
\noindent We have argued that the quantum gravitational world is structured by at least three fundamental laws: a maximal velocity principle, an universal bound for the mass-energy that one may store in a given region of spacetime, and the quantum of action whose physical interpretation is less clear. These inequalities require some specific combinations of the fundamental scales $(l_p,m_p,t_p)$ to emerge. The three inequalities we propose are responsible, in practice, for the emergence of the ratios, $c=l_p/t_p$, $G/c^2=l_p/m_p$, and $\hbar/c=m_p l_p$.\\
\\
The detailed analysis of the second proposition led us to advocate a new covariant mass bound conjecture, whose relation to Penrose inequality has been stressed. We find it quite plausible that a spacetime is free of horizons if and only if for any spacelike surface $B$ in this spacetime, the integrated mass on the light-sheets $L[B]$ satisfies
$$
\frac{m[L[B]]^2}{c^4} < \frac{A[B]}{16 \pi G^2}.
$$
Such a manifestly covariant principle for the formation of horizons shall not introduce spurious Lorentz-violating effects when applied to specific situations -- in general, thought experiments aiming at probing the nature of spacetime and physical laws at the Planck's scale. We have also argued that the relativistic quantum of action (so to say), i.e. $\hbar/c$, can be given a geometrical meaning in four dimensional spacetimes, with a parent inequality:
$$
A[B] > \frac{\pi \hbar^2}{m[L[B]]^2 c^2},
$$
whose generalization to $D \neq 4$ spacetimes leads to the provocative idea that $\hbar$ might be dimension-dependent. These two formulas, in a quantum context, shall presumably be understood as the expectation value for the area, considered as an operator.\\
\\
We have commented on the fate of Lorentz symmetry near the Planck scale, which, to the author's opinion, is still open. We also presented a simple argument showing that waves with subplanckian wavelength are not \textit{a priori} excluded by the first principles, and this seems to contradict several phenomenological approaches to quantum gravity, in particular DSR theories (at least the subset of those theories where the wavelength is indeed bounded by below).\\
\\
Finally, we argued that a system cannot experience an acceleration greater than the inverse of its size, a criteria that, presumably, shall also be given a covariant form by squaring this inequality, and making an area appear. Using previous results we then obtained Caianiello's maximal acceleration 
$$
 a<\frac{m c^3}{\hbar},
$$
for small mass particles, and $a<c^4/G m$ for particles of larger masses. We took profit of the great issues of modern astrophysics and cosmology to realize that Caianiello's maximal acceleration may actually give a physical origin to Milgrom's constant $a_0$. This would require an ultra-light particle that could be associated to Dark Energy. Admittedly however, toy-models shall be designed in order to make this possibility a testable theory. This is left for future work.\\
\\
On a more general note, the careful reader will have noticed that we dismissed, most of the time, the parameter $t$ which is supposed to characterize a physical system, besides its mass/energy and size. We did so because it is actually not clear to what $t$ precisely refers. Accordingly, we have given an interpretation to the constants $G/c^2$ and $\hbar/c$, but these ratios are specific in the sense that the Planck time has disappeared in their expression. All the previous considerations, and especially the two formulas for the area, are thus somewhat of static nature. Clearly one cannot give a satisfactory interpretation of the nature of $\hbar$ without a fine understanding of its meaning with respect to time evolution. How such considerations would relate to the ``static'' bounds we have found should be investigated in more depth.
\begin{acknowledgments}
J.-P. B. is FSR/COFUND postdoctoral researcher at naXys and wishes to thank A. Füzfa for encouraging support, as well as C. Bardavid, B. Famaey, F. Lechenault, and J. Larena for useful discussions.
\end{acknowledgments}

\bibliographystyle{hunsrt}
\bibliography{mybib}

\end{document}